\documentclass[a4paper]{jpconf}
\usepackage{graphicx,amsmath}
\begin{document}
\title{Microscopic Description of Induced Fission}

\author{Nicolas Schunck}

\address{Physics Division, Lawrence Livermore National Laboratory, CA 94551, USA}

\ead{schunck1@llnl.gov}

\begin{abstract}
Selected aspects of the description of neutron-induced fission in $^{240}$Pu
in the framework of the nuclear energy density functional theory at finite
temperature are presented. In particular, we discuss aspects pertaining to the
choice of thermodynamic state variables, the evolution of fission barriers as
function of the incident neutron energy, and the temperatures of the fission
fragments.
\end{abstract}

\newcommand{\gras}[1]{\boldsymbol{#1}}

\section{Introduction}

Nuclear fission remains one of the most complex physics phenomena in nature.
Powerful phenomenological models, finely tuned to a wealth of precise
experimental data, have given us a good qualitative, and at times quantitative,
understanding of the phenomenon. However, these models of fission lack a firm
connection with the theory of the nuclear force. This jeopardizes their ability
to supply the high-precision predictions currently needed to understand the
later stages of the formation of elements in supernova, or to simulate new
prototypes of nuclear reactors.

Some of the basic concepts of a microscopic theory of fission were already laid
out in the eighties by the theory group at CEA Bruy\`{e}res-le-Ch\^{a}tel
in France \cite{[Ber84],[Ber89]}. Starting from an effective interaction between
nucleons that embeds (in a phenomenological way) in-medium many-body correlations,
the nuclear self-consistent mean-field theory provides a generic framework to
construct ever more accurate representations of the nuclear wavefunction
\cite{[Ben03]}. To describe fission, a number of collective variables must be
introduced. The potential energy surface in the collective space thus defined
serves as a basis to perform a time evolution of the nuclear wave packet in the
framework of the time-dependent generator coordinate method. As evidenced from
the few applications reported, the method is very promising \cite{[Gou05]}.

However, at the time of its inception, the necessary computing power was
lacking so that this approach could not yet compete with more empirical models.
Today, the fast development of leadership class computers and interdisciplinary
collaborations such as the NUCLEI project open new perspective \cite{[unedf]}.
In addition, recent progress in deriving quality energy functionals for fission
applications \cite{[Kor12]} and of understanding the quantum mechanics of scission
\cite{[You11]} suggests the microscopic approach to fission has enough potential
to become predictive.

In this work, we discuss some aspects of the microscopic theory of fission in
the poster-child example of the neutron-induced fission of $^{240}$Pu at low
energies. After a brief reminder of the theoretical framework, we show the
evolution of the fission barriers as function of the nuclear temperature, and
discuss some of the practical aspects related to such simulations.

\section{Theoretical framework}

Our formalism is based on the nuclear energy density functional theory (DFT) with
local energy functionals of the Skyrme type \cite{[Per04]} for the particle-hole
channel, and contact density-dependent interactions with mixed volume and surface
character for the particle-particle channel \cite{[Dob02]}. The nucleus is
described at the Hartree-Fock-Bogoliubov approximation: its ground state is a
quasi-particle vacuum that depends on a number of collective variables $\gras{q} =
\{ q_{i} \}_{i=1,\dots,N}$, such as, e.g. the expectation value of multipole moments
or angular momentum. The set of the $N$ collective variables defines a point
$\gras{q}$ in the collective space. For each point in the collective space, the
requirement that the energy be minimal with respect to variations of the
generalized density leads to the HFB equations; solving the latter defines the
one-body density matrix and the pairing tensor, and thereby all observables of
interest.

The finite-temperature extension of the HFB formalism (FT-HFB) was introduced in
nuclear physics more than 30 years ago \cite{[Goo81]}. We only recall that in the
finite-temperature HFB theory, the ground-state is a statistical superposition
of pure quantum states, characterized by a density operator $\hat{D}$. Adopting
the HFB approximation for the density operator leads to the finite-temperature
HFB equations: they are formally equivalent to their form at zero-temperature,
only the expression of the density matrix and pairing tensor is different. Most
importantly, the Wick theorem still applies, meaning that all observables are
computed from the trace (in the given representation) of the corresponding
operator and the one-body density matrix.

In addition to the density operator, macroscopic thermodynamical concepts
are also used to describe the system at finite temperature. In particular,
several thermodynamical potentials are available to describe the nucleus, each
coming with its set of state variables. The two most useful are (i) the internal
energy $E(V,S,X)$, where $V$ is the (constant) nuclear volume, $S$ the entropy
and $X$ any additional extensive state variable needed to characterize the
nucleus (expectation value of multipole moment, for example) and (ii) the
Helmholtz free energy $F(V,T,X)$ with $F = E - TS$ and $T$ the temperature.
The FT-HFB equations are obtained by minimization of the grand potential at
temperature $T$, and naturally yield the free energy in the isothermal
representation (constant $T$). Passage to an isentropic description (constant
$S$) is sometimes needed, as discussed below.

\section{Induced fission and finite-temperature formalism}

In the description of induced fission in the finite-temperature DFT framework,
the first step is to compute the potential energy surface of the nucleus as a
function of the chosen collective coordinates and the nuclear temperature. In
the following, we work in a four-dimensional collective space characterized by
the expectation values of the axial, $\hat{Q}_{20}$, and triaxial, $\hat{Q}_{22}$,
quadrupole moment, mass octupole moment $\hat{Q}_{30}$ and hexadecapole moment
$\hat{Q}_{40}$. All calculations have been performed with the SkM* parametrization
of the Skyrme functional \cite{[Bar82]}. In the particle-particle channel, the
proton and neutron pairing strengths were fitted to the experimental 3-point
formula for the odd-even mass differences in $^{240}$Pu. A cutoff of
$E_{\text{cut}}=60$ MeV limits the number of quasi-particles taken into account in the
calculation of the densities.

Calculations were performed with the DFT solvers HFBTHO \cite{[Sto12]} and
HFODD \cite{[Sch11]}. In both codes the solutions to the HFB equations are
expanded in the one-center harmonic oscillator (HO) basis. In HFBTHO axial
and time-reversal symmetry are assumed so that solutions are labeled by the
projection of the angular momentum on the $z$-axis. By contrast, HFODD is
fully symmetry-unrestricted, and can, in particular, describe triaxial and
parity-breaking shapes. The two programs have been benchmarked against one
another and agree within a few eV for an axial configuration \cite{[Sto12]}.
In practice, the HFB wavefunctions were expanded onto $N_{\text{states}} =
1100$ HO basis states with contributions from up to $N_{\text{max}}=30$
oscillator shells.

\subsection{Evolution of fission barriers}

\begin{figure}[h]
\begin{center}
\includegraphics[width=30pc]{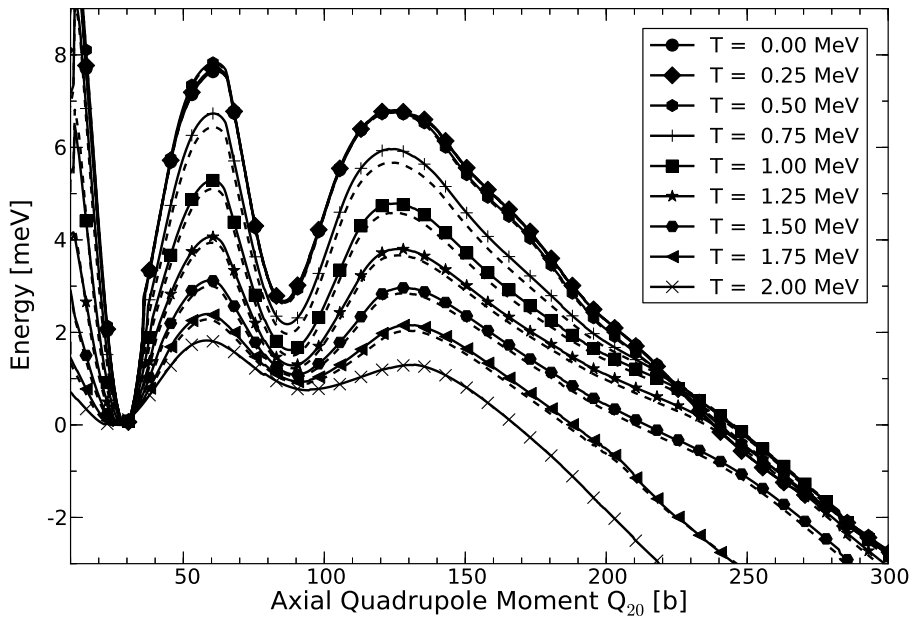}
\caption{\label{fig:maxwell} Evolution of fission barriers in $^{240}$Pu as
a function of temperature along the most probable fission path. Plain lines
curves with a symbol show the free energy at constant temperature (isothermal
process), the dashed-line curves next to them show the corresponding internal
energy at constant entropy (isentropic process). }
\end{center}
\end{figure}

In figure \ref{fig:maxwell}, we show the evolution of the free energy $F = E-TS$
as function of the expectation value $Q_{20}$ of the axial quadrupole moment. All
three other degrees of freedom are locally minimized: in practice, the first
barrier has a non-zero value of $\hat{Q}_{22}$ while the octupole moment is
non-zero past the second barrier. Note the gradual disappearance of the barriers
as the temperatures increases.

In applications of induced fission, the primary motivation has to do with the
properties of the fission fragments rather than those of the fissioning compound
nucleus. Of particular importance are the charge and mass distributions of the
fragments, their total kinetic energy (TKE), and their excitation energy. The
latter is a major input in the reaction codes that model the neutron and gamma
spectrum generated during fission. Estimates of TKE and excitation energy require,
in principle, to work in the isentropic representation. The Maxwell relations of
thermodynamics state that, for any (extensive) collective coordinate $q_{i}$,
\begin{equation}
\left. \frac{\partial F}{\partial q_{i}}\right|_{T,q_{j\neq i}}
=
\left. \frac{\partial E}{\partial q_{i}}\right|_{S,q_{j\neq i}}.
\end{equation}
Therefore, variations of internal energy with respect to deformation at
constant entropy are equal to the variations of free energy at constant
temperature \cite{[Pei09]}.

From the set of curves $\{ F(\gras{q},T_{k}) \}_{k}$ obtained directly from the
FT-HFB calculations, one can, by numerical interpolation, reconstruct the
$E(\gras{q},S)$ for any value $S$ of the entropy. The procedure is based on a
standard spline interpolation. Finite-temperature HFB calculations produce at
each point $\gras{q}$ in the collective space the quantities $E(T)|_{\gras{q}}$
and $S(T)|_{\gras{q}}$, where from we easily obtain $E(S)|_{\gras{q}}$ for any
desired value of $S$. When normalized to the same deformation point, the curves
$E(\gras{q})|_{S}$ and $F(\gras{q})|_{T}$ should be strictly equivalent
within the interpolation errors. This is verified in figure \ref{fig:maxwell},
where all dashed lines are obtained for the values of entropy at the top of the
first barrier.

\subsection{Temperature of the fragments}

Until now, macroscopic-microscopic approaches assumed either the same temperature
in the two fragments \cite{[Wil76]}, or a ratio of temperatures proportional to
the ratio of the fragment masses $T_{1}/T_{2} \approx A_{1}/A_{2}$ \cite{[Sch10a]}.
The validity of the Maxwell thermodynamical relations in a practical case suggests
a method to predict microscopically the nuclear temperature of each individual
fragment after scission.

\begin{figure}[h]
\begin{center}
\includegraphics[width=30pc]{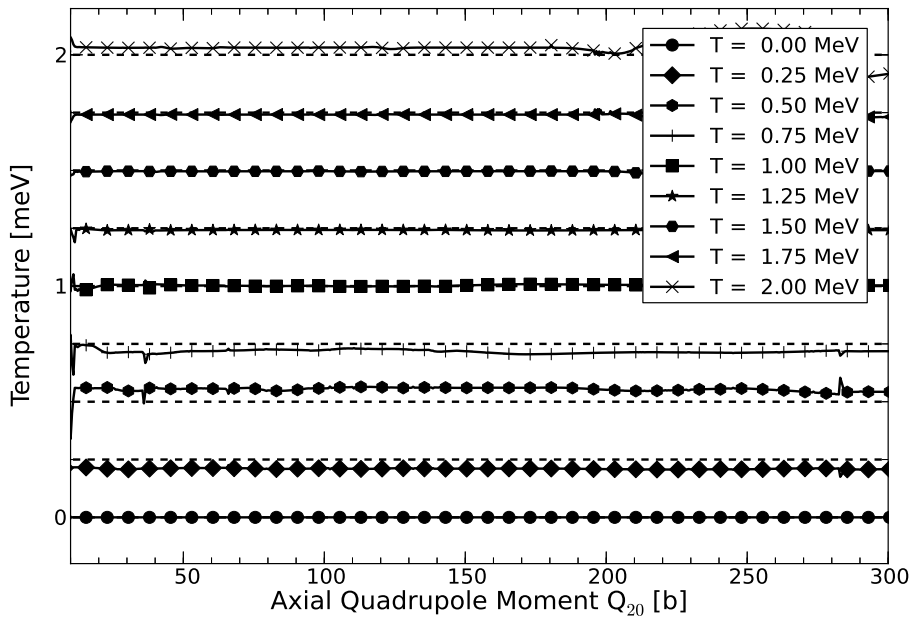}
\caption{\label{fig:temp} Temperature along the most probable fission path of
$^{240}$Pu as extracted from the entropy as function of the internal energy. }
\end{center}
\end{figure}

As seen in the previous section, the isothermal and isentropic representations
are equivalent. The advantage of the latter is that it involves only extensive
variables, and is, therefore, more amenable to the description of the splitting
of the compound nucleus into two subsystems. The scission of the compound nucleus,
characterized with an internal energy $E$ and an entropy $S$, yields two fragments,
with respective internal energy and entropy $(E_{1}, S_{1})$ and $(E_{2},S_{2})$.
Since the entropy is an extensive variable, we have $S = S_{1} + S_{2}$ (here, we
have to assume implicitly that the interaction energy between the two fragments is
`small enough'). The next step is to take advantage of the thermodynamical relation
\begin{equation}
\left. \frac{\partial S}{\partial E}\right|_{q_{i}}
=
\frac{1}{T}.
\end{equation}
For each fragment $n=1,2$, we can perform a separate FT-HFB calculation constrained
on the deformations $q_{i}^{(n)}$ of the fragment at scission. This gives us the
functions $S^{(n)}(E^{(n)})$. Taking the derivative with respect to $E^{(n)}$ gives
the temperature $T^{(n)}$ of the fragment $n$. While the value of both the internal
energy and entropy in this separate FT-HFB calculation should be the same as the
ones extracted from the compound nucleus (again: if the interaction energy
can be neglected), the temperature may be different, as it is related to the
derivative $\partial S/ \partial E$.

The method just outlined will be applied to estimating the temperatures of the
fragments in the induced fission of $^{240}$Pu in a forthcoming publication. As a
preliminary step, we estimate in figure 2 its numerical accuracy by comparing the
exact value of the temperature, as set in the FT-HFB calculation, with the numerical
estimates obtained from the $S(E)$ function. Overall, the accuracy is of the order
of 50 keV at worse, owing to interpolation errors.

\section{Conclusions}

Recent advances in the nuclear energy density functional theory, combined with
the constant increase of computing power, have enabled significant progress
toward the development of a microscopic theory of nuclear fission. In this
short article, we have presented early results related to the evolution of
fission barriers with nuclear temperature and the equivalence between isentropic
and isothermal representations.

\ack

\bigskip

Enlightening discussions with W. Younes and D. Gogny are very warmly acknowledged.
This work was supported by the U.S. Department of Energy under Contract
Nos.\ DE-AC52-07NA27344 (Lawrence Livermore National Laboratory) and
DE-SC0008499 (NUCLEI SciDAC Collaboration). An award of computer time was
provided by the Innovative and Novel Computational Impact on Theory and
Experiment (INCITE) program. This research used resources of the Oak Ridge
Leadership Computing Facility located in the Oak Ridge National Laboratory,
which is supported by the Office of Science of the Department of Energy under
Contract DE-AC05-00OR22725. It was also supported by the National Energy
Research Scientific Computing Center supported by the Office of Science of the
US Department of Energy under Contract No. DE-AC02-05CH11231.

\section*{References}
\bibliographystyle{unsrt}

\end{document}